\documentclass[12pt]{article}

\begin{document}
\title {Economics from a Physicist's point of view: Stochastic Dynamics of Supply and Demand in a Single Market. Part I}
\bigskip
\author{Vladimir T.Granik \thanks{Department of Civil Engineering(retired),University of California
Berkeley, CA 94720 }~~ and ~ Alex Granik\thanks{Department of
Physics, University of the
Pacific,Stockton,CA.95211;~E-mail:~agranik@uop.edu}}
\date{}
\maketitle
\begin{abstract}
Proceeding from the concept of rational expectations, a new
dynamic model of supply and demand in a single market with one
supplier, one buyer, and one kind of commodity is developed.
Unlike the cob-web dynamic theories with adaptive expectations
that are made up of deterministic difference equations, the new
model is cast in the form of stochastic differential equations.
The stochasticity is due to random disturbances ("input") to
endogenous variables. The disturbances are assumed to be
stationary to the second order with zero means and given
covariance functions. Two particular versions of the model with
different endogenous variables are considered. The first version
involves supply, demand, and price. In the second version the
stock of commodity is added. Covariance functions and variances of
the endogenous variables ("output") are obtained in terms of the
spectral theory of stochastic stationary processes. The impact of
both deterministic parameters of the model and the random input on
the stochastic output is analyzed and new conditions of chaotic
instability are found. If these conditions are met, the endogenous
variables undergo unbounded chaotic oscillations. As a result, the
market that would be stable if undisturbed loses stability and
collapses. This phenomenon cannot be discovered even in principle
in terms of any cobweb deterministic model.
\end{abstract}
\begin{verse}
 \footnotesize{Only the simple theories that
can be explained\\
to an intelligent outsider (one's wife)\\
turn out to hold up in economics.}

                                         P. A. Samuelson
\end{verse}

\section{Introduction}

It is universally accepted wisdom that supply and demand
interaction is the main driving force of a market economy. The
word "interaction" means that supply and demand, as well as the
price they depend upon, change with time and are therefore not
static phenomena but dynamic processes. These price-quantity
processes can develop in different ways. They can converge to an
equilibrium state, oscillate, collapse or explode, all in a
regular or chaotic manner. To gain insight into this complex
behavior, supply-and-demand problems should be treated not only
statically, as in economics textbooks, but also dynamically.\\

Yet surprisingly, the dynamic consideration of these problems has,
almost completely, given way to a static analysis. As a result,
the inadequate static approach has led to a limited and sometimes
even incorrect understanding of the price-quantity interaction. A
case in point is the single one-commodity market with one seller
and one buyer. There are some quasi-dynamic models, to be briefly
discussed below, expressly designed for a study of stability of
supply-and-demand equilibrium price in such a market.\\

No wonder that such specific models fail to deal with the whole
interactive process of supply, demand, and price. Instead, this
process is routinely discussed from the static point of view based
on the well-known oversimplified assumptions:

1. Supply $S$ and demand  $D$ are functions only of a price P and
do not depend on time $t$. The variables $S, D, P$, which are
included in the analysis, constitute an endogenous set (a market)
${\mathbf{E}}=\{S,D,P\}$.\

2. The plots of  $S(P)$  and  $D(P)$  have respectively positive
("upward") and negative ("downward") slopes everywhere. These
curves  intersect at a point of equilibrium
${\bf{E^e}}=\{S^e,D^e,P^e\}\subset {\bf E}$   where supply equals
 demand ($S^e=D^e$) and the equilibrium price $P^e$
clears the market.\\

Suppose that a set $\bf G$  of all $exogenous$ variables which can
disturb the market equilibrium is excluded from the analysis. This
is equivalent to an assumption that ${\bf G}=\emptyset$  where
$\emptyset$ is an empty set. Yet there still may be an endogenous
force ,say, the seller who raises the price $P$ that can displace
the market $\bf E$ from the equilibrium point $\bf E^e$  to a
non-equilibrium state ${\bf E^n}=\{S^n,D^n,P^n\}\neq {\bf E^e}.$
In the end, however, as the static analysis shows, the market $\bf
E$ will automatically return from $\bf E^n$ to $\bf E^e$ . Hence,
according to the static consideration, the equilibrium of the
simplest market ${\bf E}=\{S,D,P\}$ is always stable provided
${\bf G }=\emptyset$\\

This discovery stands so high in economics that it is frequently
called the "law" of supply and demand (e.g., \cite{BB}). Even its
introductory part,-the existence of an equilibrium price $P^e$,
the fact already known in the 18-th century,-was praised by Thomas
R. Malthus as the "first, greatest, and most universal principle"
of economics. To a certain extent this is true since the "law" of
supply and demand can not only give us  information about  the
equilibrium  state $\bf E^e$ , but also predict in some instances
the impact of the exogenous variables ${\bf G}\neq \emptyset$ on
the market ${\bf E}=\{S,D,P\}$\\

For example, a static analysis of an aggregate one-commodity
market (macroeconomics) enabled one to gain an initial insight
into the enigmatic phenomena of unemployment, inflation, etc., and
explain how they appear when either the aggregate supply ($AS$) or
the aggregate demand ($AD$) curve shifts in response to the
variations in the exogenous set $\bf G$ . Despite, however, its
merits, the famous "law" of supply and demand has to be taken with
a grain of salt because actually it is not a universal principle.\

Firstly, the change in the exogenous variables $\bf G$  often
leads to the "shifts on both the demand and supply side at the
same time, so the simple rules [the above explanations] ... don't
work" \cite{SN}. What happens in this case to, say,
macroeconomics, or a market of commodities, depends on the shifts'
magnitude which the famous "law" cannot predict.\

Secondly, contrary to the conventional assumption, the $AS$ and
$AD$ static curves can be "perverse." This means that "$AS$, at
least in some situations, slope downward and/or $AD$ may slope
upward" and "the usual implications of the macro $AS-AD$ analysis
may be misleading." \cite{P}\

Thirdly, and most importantly, even if the set of exogenous
variables $\bf{G}\neq \emptyset$ the static "law" of supply and
demand does not work either as long as it disregards the
time-dependent nature of the endogenous variables $\bf E$ .
Indeed, let us instead of the familiar static set
${\mathbf{E}}=\{S,D,P\}$ consider a new dynamic set
${\mathbf{E}(t)}=\{S(t),D(t),P(t)\}$ where the endogenous
variables $S, D, P$ (supply, demand, and price) are functions of
time $t$.\\

Now, since all the three variables change with time, so will do
both the curves of supply and demand as functions of the price.
Consequently, at different points the curves will have, generally
speaking, different slopes which can be positive, or negative, or
zero. These dynamic phenomena undermine the fundamental feature of
the static "law" of supply and demand,-the fixed positive
(negative) slope of the supply (demand) curve,-and hence
invalidate the "law's" assertion that the market equilibrium is
always stable if the exogenous variables $\bf {G}\neq \emptyset.$
Because of this fault, the static "law" becomes impractical and
can perhaps serve, as it does, only as a first step in
understanding supply-and-demand problems.\\

Such "myopic preoccupations of traditional equilibrium analysis"
\cite{S} persisted for a long time. Only the last few decades have
seen a monotonic increase in studies of price-quantity dynamics.
The respective dynamic models have been developed on the basis of
(1) differential and (2) difference (cobweb) time-dependent
equations.\

The first ("differential") direction of research goes back to
Hicks \cite{H} and Samuelson \cite{PS}. They focused on the study
of stability of supply-and-demand equilibrium in a single and
multiple markets in terms of differential equations. Because of
such a narrow objective, Hicks' and Samuelson's models were too
limiting. They were incapable of dealing with the general problem
of price-quantity evolution although the time-dependent
differential equations did make it possible. Consequently, by a
strange coincidence, the "differential" direction in
price-quantity dynamics gave way to the cobweb models. Such models
were introduced in the 1930's (e.g., \cite{E}) and have since then
became the prevailing theoretical tools for studying
price-quantity dynamics.\\

One of the first who employed a special cobweb model for analyzing
the price stability was Samuelson \cite{PS}. His model came to be
known as a naive expectations theory. Later on a more
sophisticated (and up to now widely used) cobweb model was
proposed based on a concept of adaptive expectations. The model is
reviewed in detail in our forthcoming paper. Now we only point out
some of the shortcomings of the adaptive expectations theory (see,
e.g., \cite{Ho}).\

1. In this theory, the curves of supply $f_S$ and demand $f_D$ are
supposed to be "rigid," their shape is fixed and time-independent.
This brings us back to the simplistic static assumption adopted in
textbooks. But in price-quantity dynamics there are no "rigid"
supply and demand curves. As has been already mentioned, both the
curves are generally dependent on and changing with time.\

2. The governing equations of the adaptive expectations theory are
explicitly deterministic because no  stochastic components are
included. Only by a special choice of the key deterministic
functions $f_S$ and $f_D$ , the cobweb model may be able to reveal
some stochastic features of the price-quantity dynamics. But
because there are in fact no "rigid" functions $f_S$ and $f_D$,
this particular approach is in general overly restrictive.
Therefore a better way to study the chaotic behavior of the
price-quantity process is to address it directly, by incorporating
stochastic functions into governing equations. Now we can proceed
with our new theory.

\section{ The Stochastic Dynamic Model}

We begin with two assumptions.\\

1. It is clear that a profit-minded supplier should acutely aware
of his/her costs. He/she will therefore control the actual output
of commodity $S(t)$  in view of the difference $P_{St}=P(t)-P_S$
between the market price $P(t)$   and some characteristic price
$P_S$ which includes all costs and the desired profit. It is
reasonable to believe that the actual supply $S(t)$  will
in-crease if the net price $P_{St}> 0$  , or decrease if $P_{St} <
0$  . So, $P_S$ can be interpreted as the seller's $borderline$
price above which the supply of commodity will rise or below which
it will fall.\\

2. It is also clear that a sensible buyer will adjust his/her
actual demand for commodity $D_t$  based on the difference
$P_{Dt}=P(t)-P_D$ between the market price $P(t)$  and a
characteristic price $P_D$  . The latter is formed by the buyer's
needs and financial opportunities, as well as by his/her tastes
and other unspecified psychological, physiological, etc. factors.
The higher the price $P_D$ , the higher is the buyer's willingness
to purchase the commodity. In other words, the actual demand
$D(t)$ is likely to increase if the net price $P_{Dt}<0$ , or
decrease if $P_{Dt}>0$. Hence, $P_D$ can be viewed as the buyer's
$borderline$ price below which the demand for commodity $D(t)$
will rise or above which it will fall.\\

All the factors influencing the price $P(D)$,-the buyer's needs,
financial opportunities, tastes, etc.,-are, to a different degree,
random by nature. We may therefore express  $P(D)$ as a sum of two
components $P(D)=\langle P_D\rangle+\overline P_D$ where $\langle
P_D\rangle \neq 0$ is a deterministic part (a mean) of $P_D$ while
$\overline P_D$ is a stochastic disturbance of $P_D$  with a zero
mean $\langle {\overline P}_D\rangle =0.$ Besides, the seller's
borderline price $P_S$ is similar to $P_D$ and is therefore also
stochastic. Yet in the present paper, for the sake of simplicity,
the price is taken to be a deterministic quantity. \\

Applying these two assumptions (as well as some others) to a
rather inclusive rational expectations model (e.g., \cite{Farm}),
we have derived governing relations of our dynamic theory:

\renewcommand{\theequation}{\thesection.\arabic{equation}}
\begin{equation}\label{eq:2.1}
\left \{ \begin{array}{c}
\dot{S}(t)\\\dot{D}(t)\\\dot{P}(t)\end{array} \right \}= \left
\{\begin{array}{cr}
a[P(t)-P_S]+k[D(t)-S(t)]\\
b[<P_D>-P(t)]\\
c[D(t)-S(t)]\end{array} \right \}+ \left \{\begin{array}{cr}
0\\
b\overline{P}_D\\
0 \end{array}\right \}+ \left \{\begin{array}{cr}
\phi_S(t)\\
\phi_D(t)\\
\phi_P(t) \end{array}\right \}
\end{equation}\\

It is a system of three stochastic first-order linear differential
equations for the set of three endogenous variables ${\bf
E}(t)=\{S(t),D(t),P(t)\}.$ In Eqs. (\ref{eq:2.1}), dots denote
differentiation with respect to time $t$; the quantities $a,b,c,k$
are non-negative constants; $\phi_S(t),\phi_D(t),\phi_P(t)$
represent exogenous deterministic functions influencing supply,
demand, and price.\\

Now a natural question arises: What is the economic interpretation
of the governing equations (\ref{eq:2.1}). Let us first look at
the first of these equations which we designate as
${(\ref{eq:2.1})}_1$. Its meaning seems quite clear:\

(i) If the market price exceeds the seller's borderline price,
supply will rise. If the seller's borderline price exceeds the
market price, supply will fall.\

(ii) If demand exceeds supply, supply will rise. If supply exceeds
demand, demand will fall. The meaning of next equation designated
as ${(\ref{eq:2.1})}_2$is no less clear:\

(iii) If the buyer's borderline price exceeds the market price,
demand will rise. If the market price exceeds the buyer's
borderline price, demand will fall.\

Lastly, let us the third of Eqs.(\ref{eq:2.1}) designated as
${(\ref{eq:2.1})}_3$. Its meaning has been given by Samuelson's
dictum \cite{PS}:\

(iv) "If at any price demand exceeds supply, price will rise. If
supply exceeds demand, price will fall."\

We thus see that the paraphrase of Eqs. (\ref{eq:2.1}) is very
simple. If we had started from this literal interpretation (that
is, by inverting our approach), Eqs. (\ref{eq:2.1}) could have
easily been written $tout de suite.$ Yet we have avoided this path
and derived the dynamic model (\ref{eq:2.1}), as mentioned before,
differently,- in terms of the more fundamental rational
expectations model. This has been done on purpose, in order to
show that our model (\ref{eq:2.1}) not only differs from the
rational expectations model, but also has a certain affinity with
it which would otherwise have been difficult, if not impossible,
to see.\

There is also the other side of the coin. Once the stochastic
equations (\ref{eq:2.1}) are obtained, they lend themselves well
to developing more inclusive dynamic models of supply-and-demand
to be dealt with elsewhere. It is important to note that unlike
Eqs. (\ref{eq:2.1}) and without them such advanced models are not
directly derivable from the rational expectations model (cf.
\cite{Farm}).

\section{A Closed Deterministic Market}

 First we consider a closed deterministic market,
that is, a market without exogenous variables and stochastic
disturbances. This means that in the right-hand side of the model
(\ref{eq:2.1}) we should omit the second and the third vector
components and thus obtain \setcounter{equation}{0}
\renewcommand{\theequation}{\thesection.\arabic{equation}}
       \begin{equation}\label{eq:3.1}
 \left\{\begin{array}{c}
\dot{S}(t)\\
\dot{D}(t)\\
\dot{P}(t)
\end{array}\right\}=
\left\{\begin{array}{cr}
a[P(t)-P_S]+k[D(t)-S(t)]\\
b[<P_D>-P(t)]\\
c[D(t)-S(t)]
\end{array}\right\}
\end{equation}

It is a system of three deterministic first-order linear
differential equations which can easily be solved by using any
traditional techniques (see, e.g., \cite{Dit},\cite{John}).
Analytic solutions of Eqs. (\ref{eq:3.1}) are simple but rather
cumbersome. Therefore,in what follows, we restrict ourselves to
the illustrations of typical results.
\begin{center}
C a s e   1
\end{center}
Suppose that there is no damping, that is, the constant k = 0. In
this particular case, the solution of (\ref{eq:3.1}) is
comparatively compact:
\begin{equation}\label{eq:3.2}
S(t)=\frac{a}{a+b}[S(0)-D(0)]cos(\beta
t)+\frac{a}{\beta}[P(0)-A_1]sin(\beta t)+A_2t+A_3
\end{equation}
\begin{equation}\label{eq:3.3}
D(t)=-\frac{b}{a+b}[S(0)-D(0)]cos(\beta
t)+\frac{b}{\beta}[P(0)-A_1]sin(\beta t)+A_2t+A_3
\end{equation}
\begin{equation}\label{eq:3.4}
P(t)=[P(0)-A_1]cos(\beta t)-\frac{\beta}{a+b}[S(0)-D(0)]sin(\beta
t)+A_1
\end{equation}

where $S(0),D(0),P(0)$   stand for initial values of supply,
demand, and price, and the parameters $A_i(i=1,2,3),\beta$   are
defined as
\begin{equation}\label{eq:3.5}
\left\{\begin{array}{c}
A_1)\\
A_2\\
A_3\\
\beta
\end{array}\right\}=
\left\{\begin{array}{cr}
(aP_S+b\langle P_D\rangle)(a+b)^{-1}\\
ab(<P_D>-P_s)(a+b)^{-1}\\
\lbrack aD(0)+b S(0)\rbrack (a+b)^{-1}\\
 \sqrt{(a+b)c}
\end{array}\right\}
\end{equation}

We see that according to Eqs. (\ref{eq:3.2})-(\ref{eq:3.5}),\

$\bullet$ If $\langle P_D \rangle > P_S$ , then $A_2>0$.
Consequently, both supply and demand increase in time,and the
market is booming.\

$\bullet$ If in addition $S(0)\neq D(0)$ and/or $P(0) \neq A_1$ ,
then the monotonic increase of supply and demand is accompanied by
undamped oscillations. \

$\bullet$ If $\langle P_D \rangle < P_S$ , then $A_2< 0$. As a
result, both supply and demand decrease in time and the market
goes south. \

$\bullet$ If simultaneously $S(0)\neq D(0)$ and/or $P(0) \neq A_1$
, then the monotonic decrease of supply and demand is followed by
undamped oscillations.\

$\bullet$ If  $\langle P_D \rangle = P_S$ , then $A_2= 0$.
Accordingly, both supply and demand either oscillate [when
$S(0)\neq D(0)$ and/or $P(0) \neq A_1$   ] or remain equal to
$A_3= constant.$\

$\bullet$ If $P(0) \neq A_1$ and/or $S(0)\neq D(0)$, then the
price $P(t)$ oscillates.\

$\bullet$ If both $P(0)= A_1$ and $S(0)= D(0)$, then the price
$p(t)$ does not change and remains equal to the initial value
$P(0)=A_1$ .\\

 The foregoing brief analysis leads to important conclusions.\

   1. The supply-and-demand dynamics is mainly influenced by the ratio
   $\alpha=\langle P_D\rangle/P_S$  .
   This ratio defines what may be called the market asymmetry.\

   2. If $\alpha = 1$, the market is symmetric since both the seller and
   the buyer adhere to the same borderline price $\langle P_D\rangle=P_S.$
    As a result, the market will neither expand nor collapse, i.e., it is
    in a sense stable.\

    3. If $\alpha \neq 1$, the market is asymmetric because the seller and
    the buyer adhere to different borderline prices. Consequently, the
    asymmetric market is, in the same sense, unstable, -it will either boom
    or collapse.\\

The above new criteria of stability are drastically different from
the corresponding stability conditions established by such figures
as L. Walras, A. Marshall, J.R. Hicks, and P.A. Samuelson
\cite{PS1}. The new criteria also show that, contrary to the
static "law" of supply and demand, the equilibrium of a single
market is not always stable even if the set of endogenous
variables is empty [${\bf G}(t)=\emptyset$] as has already been
mentioned in Introduction.\\

It is also interesting to observe that the seller's borderline
price $P_S$  is usually hidden from the buyer and, vice versa, the
buyer's borderline price  $\langle P_D\rangle$ is generally
unknown to the seller. The obvious inference is that incomplete
(asymmetric) information about the characteristic prices $P_S$ and
$\langle P_D\rangle$ available respectively to the seller and to
the buyer may have either a beneficial or an adverse effect on the
market. This phenomenon has a direct bearing on a theory of
markets with asymmetric information \cite{Ak}.\

Particular examples of the closed deterministic market described
by (\ref{eq:3.1}) will be considered in the next paper.

\end{document}